\documentstyle[rotate,aaspptwo]{article}
\def\Xray{\hbox{X-ray}}
\def\gray{\hbox{$\gamma$-ray}}
\def\HI{\ion{H}{1}}
\def\HeI{\ion{He}{1}}
\def\HeII{\ion{He}{2}}
\def\NaI{\ion{Na}{1}}
\def\MgII{\ion{Mg}{2}}
\def\FeII{\ion{Fe}{2}}
\let\simgt\gtrsim

\makeatletter
\def\fps@figure{tbp}
\def\fps@table{tbp}
\makeatother

\def\doublebox!#1!#2!{\hbox{$\vcenter{%
  \tabskip0em\halign{\strut##\hfil\cr #1\cr #2\cr}}$}}
\def\tablevspace#1{\noalign{\vskip #1}}
\newdimen\myskip\myskip1mm
\makeatletter
\def\thebibliography{\subsection*{REFERENCES}
\list{}{\labelwidth1.2em\leftmargin\labelwidth\labelsep\z@\parsep\z@
\itemsep\z@\itemindent-1.2em\usecounter{enumi}}
\def\refpar{\relax}
\def\newblock{\hskip .11em plus .33em minus .07em}
\sloppy\clubpenalty4000\widowpenalty4000
\sfcode`\.=1000\relax}
\makeatother
\begin{document}
\title{A Luminous Companion to SGR~1806$-$20\footnote{%
This is a preprint of a paper submitted to {\em The Astrophysical Journal
(Letters)}. No bibliographic reference should be made to this preprint.
Permission to cite material in this paper must be received from the
authors.}}
\author{M. H. van Kerkwijk, S. R. Kulkarni, K. Matthews, G.\ Neugebauer}
\affil{Palomar Observatory, California Institute of Technology 105-24,
       Pasadena, CA 91125, USA}
\begin{abstract}
We have obtained infrared spectra of the star suggested to be the
counterpart of the soft \gray\ repeater (SGR) 1806$-$20.  We found
strong emission lines similar to those seen in the spectra of the rare
Luminous Blue Variables and B[e] stars.  A \HeI\ absorption line is
also seen, from which we infer a spectral type O9--B2.  This
classification, in combination with the minimum distance of
$\simgt6\,$kpc inferred from its extinction, makes the star one of the
most luminous in the Galaxy.  We infer that it is a companion to
SGR$\,$1806$-$20, and suggest that the presence of a companion is
somehow related to the SGR phenomenon.
\end{abstract}


\section{Introduction}
Soft \gray\ repeaters (SGR) are distinguished from the classical
\gray\ bursters by their recurrence and \gray\ spectra (Higdon \&
Lingenfelter \cite{HL90}; Norris et al.\ \cite{N91}).  Of the three
SGRs known, one, SGR$\,$1806$-$20, is associated with the
center-filled, non-thermal radio nebula G\,10.0$-$0.3 (Kulkarni et
al.\ \cite{K94}), which has an \Xray\ source, AX$\,$1805.7$-$2025,
coincident with its peak (Murakami et al.\ \cite{M94}; Cooke
\cite{C93}).  The similarity to what is observed for plerions
(pulsar-powered supernova remnants) like the Crab nebula, has led to
the hypothesis that SGRs are young pulsars.  However, it is unknown
why only some pulsars become SGRs, and how a pulsar can produce the
brilliant bursts of $\gamma$~rays that are observed.  Models involving
a companion could address both issues.

In fact, a highly reddened, luminous star (star~A) has been suggested
to be the stellar counterpart of SGR$\,$1806$-$20 (Kulkarni et al.\
\cite{K95}, hereafter K95; Vasisht, Frail, \& Kulkarni \cite{VFK95}).
In this paper, we present spectroscopic observations of this star.

\section{Observations}
A K-band spectrum of star~A was kindly taken for us in August, 1994,
by our colleagues L.J.~Smith and P.A.~Crowther, using the Cooled
Grating Spectrograph~4 attached to the United Kingdown Infrared
Telescope (UKIRT) on Mauna Kea, Hawaii.  The spectrum covers
2.02--2.22$\,\micron$ with a spectral resolution
$\lambda/\Delta\lambda$ of $\sim\!700$.  Atmospheric and instrumental
effects were removed by division by a spectrum of HR$\,$7377 (with its
\HI~Br$\gamma$ line artificially removed).  Flux calibration was done
by scaling the spectrum to produce the K-band magnitude from K95. The
reduced spectrum, reproduced in Figure~1, shows a number of emission
features arising from both low and high-excitation ions (Table~1).

The emission lines are reflected as wiggles in low-resolution
($\lambda/\Delta\lambda\simeq75$) grism spectra taken earlier with the
Hale 200-inch telescope on Palomar Mountain.  These H and K-band
spectra were used by K95 to argue that star~A was not a late-type
luminous star.  New J and H-band spectra were obtained in October,
1994, with the same instrumental setup that was used by K95, except
that a new 256$\times$256 InSb array was used, and that the slit was
narrowed to 0\farcs5 to double the spectral resolution.  All spectra
were corrected for telluric and instrumental features using spectra of
HR$\,$7643, and flux-calibrated by scaling to the observed magnitudes
{}from K95.  The spectra, reproduced in Figure~2, show a number of
additional emission features (Table~1).

\begin{figure}
\rotate[r]{\centering\leavevmode\epsfysize\hsize
\epsfbox{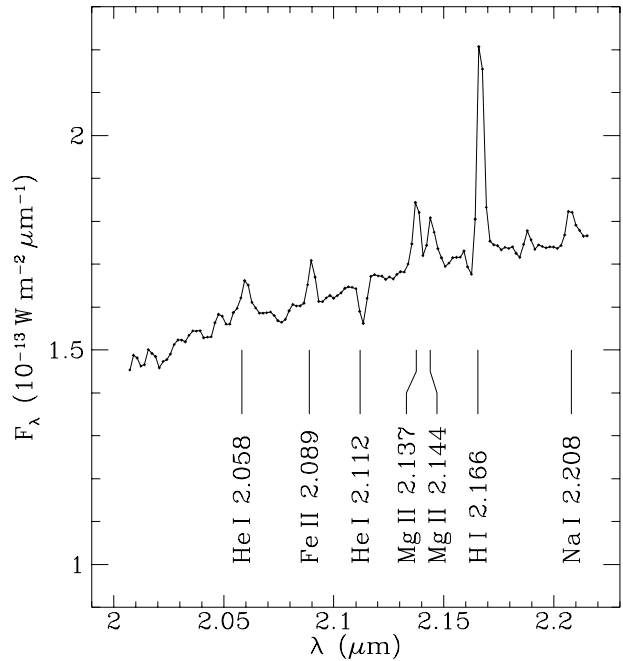}}
\caption[]{The 2.02--2.22 micron UKIRT spectrum obtained on 19 August
1994, and suggested identifications of the emission and absorption
features.  The wiggles seen around and shortward of
\HeI$\,\lambda$2.058 are due to residual telluric CO$_2$ features.  A
possibly significant feature with a P~Cygni profile is present at
2.188$\,\micron$.  If real, it could be due to \HeII$\,\lambda2.189$.}
\end{figure}

\begin{figure*}
\rotate[r]{\centering\leavevmode\epsfysize\textwidth
\epsfbox{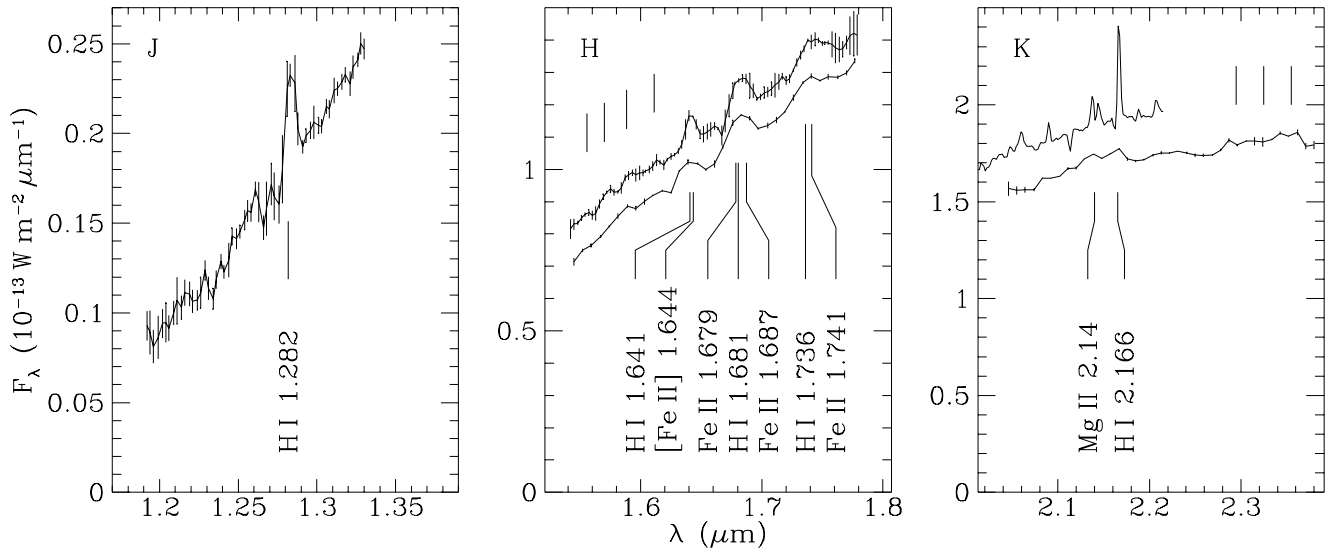}}
\caption[]{The J, H and K-band Palomar spectra.  The low-resolution
spectra in the H and K bands (lower curves) were obtained on 6 October
1993, and the higher-resolution ones in J and H (upper curve; offset
by 0.1) on 18 October 1994.  The UKIRT K-band spectrum (upper curve;
offset by 0.2), is reproduced for comparison.  Suggested
identifications of the emission features are given.  Above the H-band
spectra, also the rest wavelengths of some higher members of the
Brackett series are indicated, which seem to be in emission as well.
Above the K-band spectrum, the heads of the first-overtone CO bands
are indicated.  They may be in emission, but this needs to be
confirmed.  Possibly significant emission is also seen blueward of
Pa$\,\beta$.  This may be due partly to [\FeII]$\,$1.257.}
\end{figure*}

\begin{planotable}{lllll}
\tablewidth{0pc}
\tablecaption{Line Identifications and Equivalent Widths}
\tablehead{
\colhead{$\lambda_{\rm UKIRT}$}&
\colhead{$\lambda_{200''}$}&
\colhead{Identification}&
\colhead{$\lambda_{\rm lab}$}&
\colhead{EW} \\[.2ex]
\colhead{(\micron)}&
\colhead{(\micron)}&
\colhead{}&
\colhead{(\micron)}&
\colhead{(\AA)}}
\startdata
\nodata&1.2830(15)& \HI~$5-3$ (Pa$\,\beta$) &        1.2818&    $-22(4)$ \nl
\tablevspace{\myskip}
\nodata&1.6406(15)&\doublebox!
                    \HI~$12-4$ (Br$\,$12)!
                   [\FeII]~$a^4F_{9/2}-a^4D_{7/2}$!& \doublebox!
                                                     1.6407!
                                                     1.6436!&   $-9.4(20)$ \nl
\tablevspace{\myskip}
       &          & \FeII~$z^4F_{9/2}-c^4F_{7/2}$&   1.6787&\nl
\nodata&1.6810(20)& \HI~$11-4$ (Br$\,$11)&           1.6806&    $-15.0(20)$ \nl
       &          & \FeII~$z^4F_{9/2}-c^4F_{9/2}$&   1.6873&\nl
\tablevspace{\myskip}
\nodata&1.7401(20)&\doublebox!
                    \HI~$10-4$ (Br$\,$10)!
                    \FeII~$z^4F_{7/2}-c^4F_{7/2}$!&  \doublebox!
                                                     1.7362!
                                                     1.7414!&   $-14(3)$ \nl
\tablevspace{\myskip}
2.0591(10)&\nodata& \HeI~$2s^1S-2p^1P^0$&            2.0581&    $-3.0(10)$ \nl
\tablevspace{\myskip}
2.0893(5)&\nodata&  \FeII~$z^4F_{3/2}-c^4F_{5/2}$&   2.0888&    $-2.1(4)$ \nl
\tablevspace{\myskip}
2.1125(5)&\nodata&  \HeI~$2p^3P^0-4s^3S$&            2.1126&    $+1.8(4)$ \nl
\tablevspace{\myskip}
\doublebox!
2.1370(5)!
2.1437(5)!&2.137(4)&
                    \doublebox!
                    \MgII~$5p^2P_{3/2}-5s^2S_{1/2}$!
                    \MgII~$5p^2P_{1/2}-5s^2S_{1/2}$!& \doublebox!
                                                     2.1375!
                                                     2.1438!&\doublebox!
                                                               $-3.8(4)$!
                                                               $-2.7(4)$!\nl
\tablevspace{\myskip}
2.1660(5)\tablenotemark{a}
         &2.165(4)&
             \HI~$7-4$ (Br$\,\gamma$)&        2.1655&   $-10.8(6)$%
\tablenotemark{a} \nl
\tablevspace{\myskip}
2.2071(5)&\nodata&  \doublebox!
             \NaI~$4p^2P^0_{3/2}-4s^2S_{1/2}$!
             \NaI~$4p^2P^0_{1/2}-4s^2S_{1/2}$!&\doublebox!
                                              2.2056!
                                              2.2083!&  $-2.4(3)$
\tablenotetext{a}{%
The listed values are for the emission component.
The possible absorption component is at 2.1620(15)$\,\micron$ and
has an equivalent width of 0.7(3)$\,$\AA\\[-4ex]}
\tablecomments{%
All wavelengths are in vacuo.  EW is equivalent width.  Numbers in
parentheses indicate 90\% confidence levels in the final decimal
place.}
\end{planotable}

\section{Interpretation of the spectra}
The spectra of star~A are similar to those observed for luminous blue
variables (LBVs), such as $\eta$~Car, and B[e] stars (Allen, Jones, \&
Hyland \cite{A85}; McGregor et al.\ \cite{M88A},b; Hamann et al.\
\cite{H94}).  Similar emission lines are also observed in some young
stellar objects (YSOs), but for the two K-band spectra that we could
find, of the Becklin-Neugebauer (B-N) object (Scoville, Kleinmann, \&
Hall \cite{SKT83}) and of LkH$\alpha\,$101 (Simon \& Cassar
\cite{SC84}), the line strengths and ratios are very different from
what we observe.  In addition, for the B-N object, lines of H$_2$ are
observed, which are not present in our spectra.

One can infer from the presence of strong emission lines that there is
a substantial amount of circumstellar matter in the system.  This
could be either in the form of a disk or an outflowing wind.  In
Figure~1, Br$\,\gamma$ appears to have a blueshifted absorption
component, i.e., a P~Cygni profile.  If so, it would indicate an
outflow velocity of $\sim\!500\,$km$\,$s$^{-1}$.  The width of the
emission lines is consistent with such an outflow velocity.  It is
similar to the velocity outflow seen in $\eta$~Car (Hamann et al.\
\cite{H94}), but much larger than the outflow velocity seen in the
B-N object and LkH$\alpha\,$101 (Scoville et al.\ \cite{SKT83}; Simon
\& Cassar \cite{SC84}).

Since sodium is not expected to be neutral at the temperatures
necessary to excite helium, the simultaneous presence of both a
Na$\,$I and a He$\,$I line in the spectrum may be surprising.  It
probably indicates that there are two physically distinct
circumstellar regions, e.g., a disk and a radiation-driven wind (Allen
et al.\ \cite{A85}; McGregor et al.\ \cite{M88B}; Scoville et al.\
\cite{SKT83}).

The presence of \HeI$\;\lambda2.113$ in absorption indicates that the
star proper is hot.  From the equivalent width, one can infer a
spectral type B2--O9 (Hanson \& Conti \cite{HC94}).  For such a star,
the intrinsic IR colors are essentially neutral
($(J-H)_0=(H-K)_0\simeq0\,$mag.).  However, for some LBVs (McGregor et
al.\ \cite{M88A},b) and for all B[e] stars (Zickgraf \cite{Z90}), the
intrinsic colors are much redder ($(J-H)_0\simeq0.5\,$mag.;
$(H-K)_0\simeq1.0\,$mag.), indicating the presence of hot dust
($T\simeq1000\,$K).  We can test for the presence of dust by
dereddening the observed colors for star~A ($(J-H)=3.1\,$mag.;
$(H-K)=1.8\,$mag.; K95).  Although the interstellar extinction law is
rather uncertain, with values for $E_{J-H}/E_{H-K}$ ranging from 1.6
to 2.1 (Mathis \cite{M90}), we find that the presence of dust is
excluded.  Thus, the object is most likely not a B[e] star.

Assuming neutral colors and using an extinction curve appropriate for
the Galactic Centre (from Rieke \& Lobofsky \cite{RL85}; for this
curve, $E_{J-H}/E_{H-K}=1.7$), we infer a foreground extinction $A_V$
of $\sim\!28\,$mag.  An almost identical value is deduced from model
fitting of the \Xray\ data of AX$\,$1805.7$-$2025 (Sonobe et al.\
\cite{S94}).  K95 suggested that this extinction arises in a molecular
cloud in the direction of SGR$\,$1806$-$20, located at a distance of
about 6$\,$kpc.  If so, then given the minimum distance of 6$\,$kpc,
the absolute J magnitude of star~A would be $<\!-8.5\,$mag.  Since the
spectral type is B2 or earlier, the bolometric magnitude would be
$<\!-10.4\,$mag.\ and the bolometric luminosity $>\!10^6\,L_{\odot}$.
Only the most massive stars, including LBVs such as $\eta$~Car, have
such high luminosities (Humphreys \& Davidson \cite{HD94}).

\section{Discussion and Conclusion}
{}From the emission-line spectrum and the inferred luminosity it follows
that star~A is most likely a star similar to a LBV.  It can only be
classified as a `candidate LBV', however, since there is as yet no
evidence for the large variability that is one of the defining
characteristics of LBVs (Humphreys \& Davidson \cite{HD94}).

Extrapolating from the LMC (Humphreys \& Davidson \cite{HD79}), we
estimate that there are only a couple of hundred stars of similar
luminosity in our Galaxy.  The probability of a chance coincidence
with the radio peak (1\arcsec; Vasisht et al.\ \cite{VFK95}) or even
the quiescent \Xray\ source (11\arcsec; Cooke \cite{C93}) is
exceedingly small.  We conclude that star~A is a very luminous stellar
counterpart of SGR$\,$1806$-$20.

The impact of this new finding on our understanding of the nature of
SGRs is unclear.  If the SGR outbursts originate on or near young,
rapidly rotating neutron stars, as seems likely given the short
timescales and high luminosities (Higdon \& Lingenfelter \cite{HL90};
Norris et al.\ \cite{N91}), the neutron star cannot be an isolated
object.  If interaction with the companion causes the \gray\
outbursts, this could explain why only some young neutron stars show
bursts, while others, like the Crab, do not.

However, in the error box of the \Xray\ counterpart of SGR~0526$-$66,
associated with the supernova remnant (SNR) N49 in the LMC, there is
no star similar to a LBV or B[e] star (Rothschild, Kulkarni \&\
Lingenfelter \cite{RKL94}; Fishman, Duthie, \& Dufour \cite{FDD81}).
 From a CCD image, taken with the 2.3-m telescope at Siding Spring
Observatory in a band centered on 6620$\,$\AA\ (relatively free of SNR
emission lines), we infer a limit on the absolute magnitude
$M_{6620}>-0.7\,$mag.  Hence, any binary model for SGRs must not rely
on the presence of a massive stellar companion.

For SGR~1806$-$20, the quiescent radio (Vasisht et al.\ \cite{VFK95})
and \Xray\ (Sonobe et al.\ \cite{S94}) emission do not show the large
variations typically observed in accreting systems.  This suggests
that little interaction takes place in quiescence, perhaps because the
pulsar wind and magnetic field prevent matter from accreting.  It
could be that the quiescent emission is due to the pulsar spinning
down, and that the companion merely acts as a trigger or a catalyst
for the \gray\ bursts.  Models attributing SGR flashes to accretion of
planetesimals or asteroids (see Higdon \& Lingenfelter \cite{HL90};
Katz, Toole, \& Unruh \cite{KTU94} and references therein) may not be
as far fetched as had been thought before.

\acknowledgements We thank L.~J.~Smith and P.~A.~Crowther for
obtaining the UKIRT spectrum for us, T.~R.~Geballe for useful
discussions about the line identifications, and J.~F.~Bell and
M.~A.~Dopita for obtaining the CCD image of N49.  UKIRT is operated by
The Observatories, on behalf of the UK Particle Physics and Astronomy
Council.  M.H.v.K.\ is supported by a NASA Hubble Fellowship, S.R.K.\
by grants from the US NSF, NASA and the Packard Foundation, and
infrared astrophysics at Caltech by a grant from the NSF.


\begin{thebibliography}{}
\bibitem[1985]{A85}\reference
   Allen, D. A., Jones, T. J., \& Hyland, A. R. 1985, ApJ, 291, 280
\bibitem[1993]{C93}\reference
   Cooke, B. A. 1993, Nature, 366, 413
\bibitem[1981]{FDD81}\reference
   Fishman, G. J., Duthie, J. G., \& Dufour, R. J.  1981, Ap\&SS, 75, 135
\bibitem[1994]{H94}\reference
   Hamann, F., DePoy, D. L., Johansson, S., \& Elias, J. 1994, ApJ, 422, 626
\bibitem[1994]{HC94}\reference
   Hanson, M. M., \& Conti, P. S.  1994, ApJ, 423, L139
\bibitem[1990]{HL90}\reference
   Higdon, J. C. \&\ Lingenfelter, R. E.  1990, ARA\&A, 28, 401
\bibitem[1979]{HD79}\reference
   Humphreys, R. M., Davidson, K., 1979, ApJ, 232, 409
\bibitem[1994]{HD94}\reference
   Humphreys, R. M., Davidson, K.  1994, PASP, 106, 1025
\bibitem[1994]{KTU94}\reference
   Katz, J. I., Toole, H. A., \& Unruh, S. H. 1994, ApJ, 437, 727
\bibitem[1994]{K94}\reference
   Kulkarni, S. R., Frail, D. A., Kassim, N. E., Murakami, T.,
   \& Vasisht, G. 1994, Nature, 368, 129
\bibitem[1995]{K95}\reference
   Kulkarni, S. R., Matthews, K., Neugebauer, G., Reid, I. N.,
   van Kerkwijk, M. H., \& Vasisht, G.  1994, ApJ (in press; K95)
\bibitem[1990]{M90}\reference
   Mathis, J. S.  1990, ARA\&A, 28, 37
\bibitem[1988a]{M88A}\reference
   McGregor, P. J., Hyland, A. R., \& Hillier, D. J.  1988a, ApJ, 324, 1071
\bibitem[1988b]{M88B}\reference
   McGregor, P. J., Hillier, D. J., \& Hyland, A. R.  1988b, ApJ, 334, 639
\bibitem[1994]{M94}\reference
   Murakami, T., Tanaka, Y., Kulkarni, S. R., Ogasaka, Y., Sonobe, T.,
   Ogawara, Y., Aoki, T. \&\ Yoshida, A.  1994, Nature, 368, 127
\bibitem[1991]{N91}\reference
   Norris, J. P., Hertz, P., Wood, K. S., \& Kouveliotou, C.  1991,
   ApJ, 366, 240
\bibitem[1985]{RL85}\reference
   Rieke, G., \& Lobofsky, M.J.  1985, ApJ, 288, 618
\bibitem[1994]{RKL94}\reference
   Rothschild, R. E., Kulkarni, S. R., \& Lingenfelter, R. E.
   1994, Nature, 368, 432
\bibitem[1983]{SKT83}\reference
   Scoville, N., Kleinmann, S. G., Hall, D. N. B., \& Ridgway, S. T.
   1983, ApJ, 275, 201
\bibitem[1984]{SC84}\reference
   Simon, M., \& Cassar, L.  1984, ApJ, 283, 179
\bibitem[1994]{S94}\reference
   Sonobe, T., Murakami, T., Kulkarni, S. R., Aoki, T., \&
   Yoshida, A. 1994, ApJ, 436, L23
\bibitem[1995]{VFK95}\reference
   Vasisht, G., Frail, D. A., \& Kulkarni, S. R.  1995, ApJ (in press)
\bibitem[1990]{Z90}\reference
   Zickgraf, F.-J. 1990, in {\it Angular Momentum and Mass Loss
   for Hot Stars} (ed.\ Davidson, K., Moffat, A. F. J., \&
   Lamers, H. J. G. L. M.) 245
\end{thebibliography}
\end{document}